# Digitization and Scientific Exploitation of the Italian and Vatican Astronomical Plate Archives

Cesare Barbieri[1], Carlo Blanco[3], Beatrice Bucciarelli[4], Regina Coluzzi[5], Andrea Di Paola[5], Luciano Lanteri[4], Gian Luca Li Causi[5], Ettore Marilli[10], Piero Massimino[10], Vincenzo Mezzalira[1], Stefano Mottola[7], Roberto Nesci[6], Alessandro Omizzolo[2], Fernando Pedichini[5], Francesca Rampazzi[1], Corinne Rossi[6], Ruggero Stagni[1], Milcho Tsvetkov[8], Roberto Viotti[9]

[1]*Department of Astronomy, University of Padova*
[2]*Specola Vaticana, Castelgandolfo*
[3]*Department of Astronomy, University of Catania*
[4]*INAF Torino*
[5]*INAF Roma*
[6]*Department of Physics, La Sapienza University, Roma*
[7]*DLR Berlin, Germany*
[8] *SSADC, Bulgarian Academy of Sciences, Sofia, Bulgaria*
[9]*IASF-CNR, Roma*
[10]*INAF Catania*

**Abstract**

There is a widespread interest to digitize the precious information contained in the astronomical plate archives, both for the preservation of their content and for its fast distribution to all interested researchers in order to achieve their better scientific exploitation. This paper presents the first results of our large-scale project to digitize the archive of plates of the Italian Astronomical Observatories and of the Specola Vaticana. Similar systems, composed by commercial flat-bed retro-illuminated scanners plus dedicated personal computers and acquisition and analysis software, have been installed in all participating Institutes. Ad-hoc codes have been developed to acquire the data, to test the suitability of the machines to our scientific needs, and to reduce the digital data in order to extract the astrometric, photometric and spectroscopic content. Two more elements complete the overall project: the provision of high quality BVRI CCD sequences in selected fields with the Campo Imperatore telescopes, and the distribution of the digitized information to all interested researchers via the Web. The methods we have derived in the course of this project have been already applied successfully to plates taken by other Observatories, for instance at Byurakan and at Hamburg.

## 1 Introduction

A great amount of highly valuable information is stored in the photographic archives of many Italian Observatories and in the Specola Vaticana, with plates dating back to the end of the XIXth Century. A proper digitization of this veritable treasury is therefore of paramount importance, both for the preservation of their volatile support and for the fuller exploitation of the scientific content (see e.g. Viotti et al., 1993, Griffin, 2001). Following an encouraging pilot program (see Barbieri et al., 2002), we set up a collaboration among our Institutions on the basis of common scientific and technological interests. Details are provided in Barbieri et al. 2003a, and Barbieri et al. 2003b. Among the many possible scientific aims for exploiting digital archives, we intend to carry out the following :

- search for past transits of asteroids and comets for a better reconstruction of their orbital and physical evolution,
- discovery and inventory of high proper motion stars,
- time history variable stars in the Milky Way and external galaxies, of AGNs and QSOs,
- inventory of novae and supernovae in external galaxies,
- spectral classification over wide fields.

We present here a report of the activity carried out so far, some results and plans for the future.

# 2 The photographic archive census

The first step of our work was the inventory of the plates contained in the archives of our Institutions and the digitization of the logbooks. As shown in the following Tables, the total number of plates is very large, too large to be digitized in its entirety in a reasonable amount of time. A visual inspection of the material was therefore preliminarily done, in order to select the best material according to the priorities set by our scientific interests.

## 2.1 Asiago Observatory

The Asiago archive comprises the plates obtained by four telescopes: the 1.2 m, the 1.8 m, the 67/92 cm Schmidt and the 40/50 cm Schmidt (see Table 1). The archive is well ordered and the plates conserved in a satisfactory manner. The logbooks of all direct images and objective prism plates have been digitized, and are accessible on-line as PDF or Excel files in the Archive section of each telescope (www.pd.astro.it/Asiago). Their utilization by the international community is already very active. The logbooks of the spectroscopic observations (122 cm and 182 cm telescopes) will come in the near future

**Table 1** – The Asiago photographic archive content (77928 plates)

| Telescope | Focus and plate dimensions | Period | No. plates |
|---|---|---|---|
| **1.2 m** | | | |
| Images | Newtonian focus f/5, 9x12 cm | 1942 - 1997 | 9720 |
| Spectra | Newtonian spectrographs | 1958 - 1991 | 3220 |
| Spectra | Cassegrain spectrograph | 1951 - 1994 | 18584 |
| **1.8 m** | | | |
| Images | Cassegrain focus f/9, 12x20 cm | 1973 - 1989 | 3870 |
| Spectra | B&C, Echelle REOSC | 1973 - 1988 | 4301 |
| **S67/92 cm** | | | |
| Images | 20x20 cm | 1966 - 1998 | 16729 |
| Objective prism plates | 20x20 cm | 1966 - 1998 | 1087 |
| **S40/50 cm** | | | |
| Images (films) | Circular, 10cm diameter | 1958 - 1992 | 18411 |
| Objective prism (films) | Circular, 10cm diameter | 1958 - 1992 | 2006 |

## 2.2 Torino Observatory

The photographic material of Torino Observatory consists of approximately five thousands plates (images only, no spectra) obtained with different telescopes during a period of about 70 years, as early as 1923 (see Table 2) in Pino Torinese. This rich material is very heterogeneous in terms of its state of conservation and potential astrophysical impact; at present, it is systematically inspected and suitably archived in chemically inert envelops; an electronic log-book is in preparation.

**Table 2** – The Torino photographic archive content (~5000 plates)

| Telescope | plate dimensions | period | No. plates |
|---|---|---|---|
| 20 cm Zeiss astrograph | 18x24cm | 1923 – 1984, | 3000 |
| 38 cm Morais refractor | 20x20cm | 1971 – 1980 | 1000 |
| 105 cm REOSC reflector | 16x16cm | 1971 – 1994 | 1000 |

In addition, some 1000 plates are in store, taken with such different telescopes as GPO/ESO, JKT/La Palma, Astrograph telescope/Cape Town for different scientific purposes, from asteroids search to QSO variability. Of particular interest, given their relatively large field size and magnitude limit, we judge the series taken at the Zeiss and GPO/ESO.

## 2.3 Catania Observatory

The Catania photographic archive consists of direct images (no spectra) taken in the original location in town (Piazza Vaccarini) and more recently on the Serra La Nave (Mt. Etna station, now M.G. Fracastoro station) of the Catania Observatory. The old plates were stored in wooden boxes without any protection, and many have been lost to humidity and fungi; now, chemically inert envelops have been procured, and a restoration procedure will be attempted when possible. The plates considered worth of digitization are detailed in Table 3.

**Table 3** – The Catania photographic archive content (~2000 plates)

| Instrument and location | Period | Project, No. plates and plate dimensions | Notes |
|---|---|---|---|
| 33 cm equatorial astrograph (Piazza Vaccarini) | 1897-1907 | Astrographic Catalogue + Carte du Ciel ~500 plates (16x16 cm) | 1 |
| 33 cm equatorial astrograph (Piazza Vaccarini) | 1910 | Halley comet 32 plates (16x16 cm) | 2 |
| 33 cm equatorial astrograph (Piazza Vaccarini) | 1956-1964 | Fields of Catania astrographic zone 211 plates (16x16 cm) | 3 |
| S61/41 Schmidt Telescope (Mt. Etna) | 1968-1992 | 1140 plates (9x9 cm) | 4 |
| 33 cm photographic objective (Mt. Etna) | 1985-1992 | 67 plates (16x16 cm) | 5 |

Notes to Table 3:

1. About 1600 plates were made in the frame of the *Carte du Ciel and Astrographic Catalogue*, from 1897 to 1907 in the declination area from +47° to +54°. Unfortunately, several plates were broken or damaged due to frequent removals of Observatory properties in such a long period, and many are in very bad state of conservation. At least 500 plates are in acceptable condition for digitization. 32 plates of Comet Halley were obtained from January to June 1910. Most plates are in good conditions. A few of these plates were reproduced in the *Atlas of Comet Halley 1910 II* (Donn et al., 1986), the other plates have been rediscovered thanks to the present program. In addition, plates have been found of the great comet of 1910.

2. From 1956 to 1964, with the same astrographic equatorial, in the framework of an international astrometric program to study high proper motion stars belonging to the astrographic zone of Catania for the *Carte du Ciel*, 211 plates were obtained. Most plates are in a good state.
3. 1168 plates were obtained with the Schmidt S41/61 telescope, with a scale of 169 arcsec/mm, at the Mt. Etna station in the period from 1968 to 1992. The plates are in good conditions. Only 28 plates of this archive have not been found.
4. 67 plates of comets and stellar fields in open clusters were obtained with the 33 cm Steinheil objective of the equatorial astrograph attached to the S61/41 telescope mounting.

## 2.4 Vatican Observatory

Table 4 gives the content of the Specola Vaticana archive, with plates coming from the astrograph in its historical locations in Rome and from the Schmidt telescope in Castelgandolfo. The Vatican archive is well preserved and ordered from the very first plate. For every plate of the *Carte du Ciel* and of the *Astrographic Catalogue* there is a detailed description of the characteristics, the plate constants to do astrometry, and information concerning exposure time and weather conditions. The digitization of the logbooks of the Schmidt telescope is currently being done by the Sofia Sky Archive Data Centre, Bulgarian Academy of Sciences (http://www.skyarchive.org/SSADC).

**Table 4 –** Content of the Specola Vaticana archive (9815 plates)

| Instrument | Period | Project, No. Plates and Plate Dimensions |
|---|---|---|
| 33 cm photographic doublet | 1894-1953 | Carte du Ciel: 540 plates<br>Astrographic Catalogue : 1148 plates<br>(16x16 cm) |
| 40 cm Zeiss Refractor quadruplet | 1935-1974 | Direct images: 380 plates (30x30 cm)<br>Direct images+spectra: 3111 plates (18x24 cm)<br>Direct images: 1245 plates (13x18 cm) |
| 60 cm Zeiss Reflector | 1935-1974 | Direct images: 924 plates (9x12 cm)<br>Direct images: 145 plates (7x9 cm)<br>Direct images: 172 plates (6x9 cm) |
| 65 cm Schmidt Telescope | 1957-1986 | Direct images: 794 plates (20x20 cm)<br>Objective Prism: 1326 plates (20x20 cm)<br>Polarimeter: 30 plates (20x20 cm) |

Two examples are shown in Fig. 1.

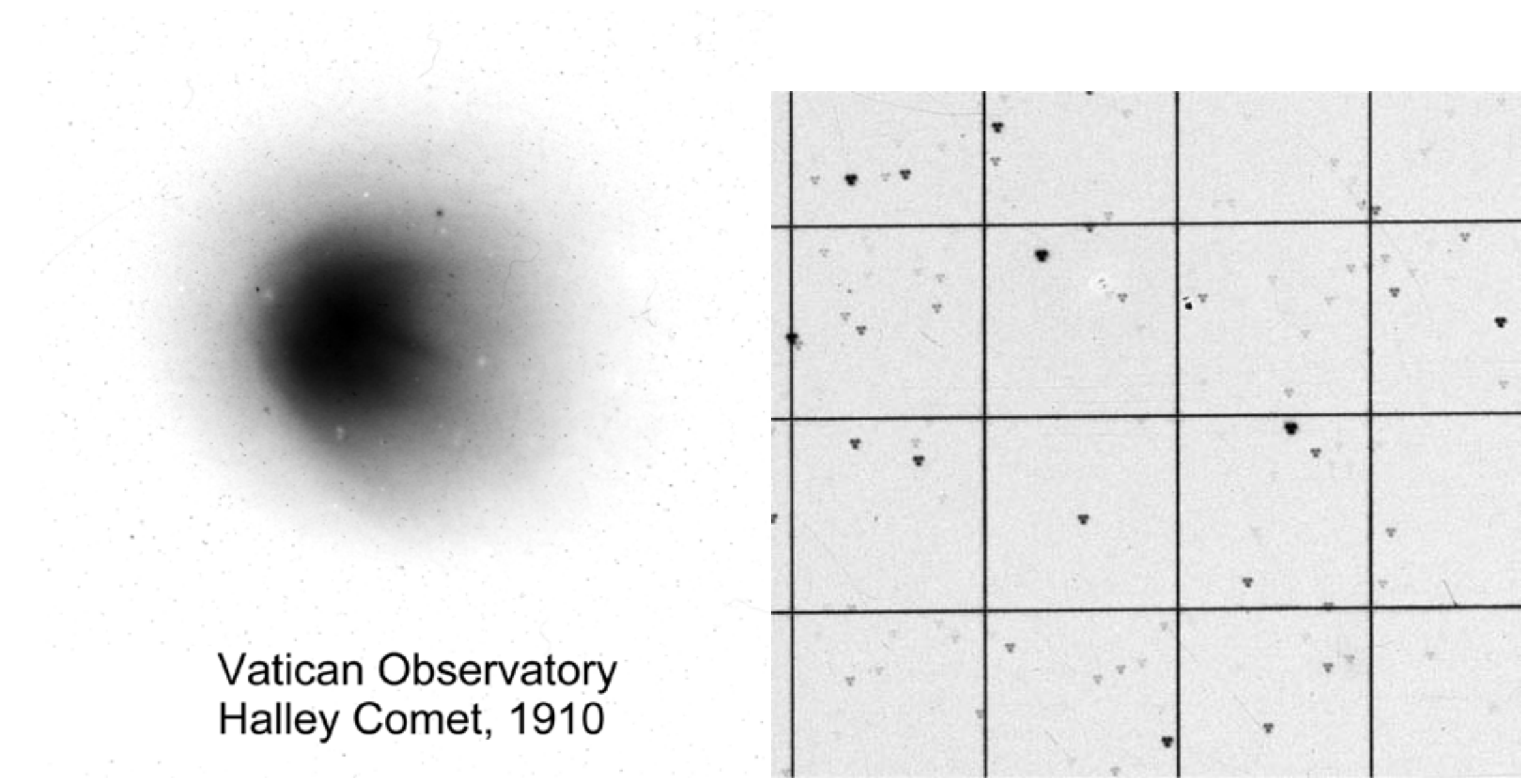

**Fig. 1 – A digitized plate of the Halley Comet (1910) and of the Carte du Ciel (plate 615, Nov. 5, 1901). No attempt has been made to clean the emulsions.**

## 3 Digitization and analysis of the plates

Several commercial flat-bed retro-illuminated scanners (A3 and A4 formats, optical resolution 1600x3200 dpi), connected to dedicated PCs, have been installed in Asiago, Padova, Catania, Roma, Castelgandolfo. The same model of scanner is used by DLR in Berlin, where a program that convincingly demonstrated the feasibility of the digitization was initially written. This program (named *Scanfits*) gradually grew to become a true user-friendly working instrument. It runs in all Windows operating systems and it provides as output a 14 or 16 bit per dot, positive or negative FITS image (including the header, where the main parameters of the scanner configuration are automatically loaded), that can be directly analyzed with IRAF, MIDAS or IDL.

Using *Scanfits*, tests have been performed on many different types of plates, both with images and spectra, to determine the spatial resolution of the scanners, and their astrometric and photometric precision, as detailed in the following paragraphs. The effective pixel size of 16 micron provides a spatial resolution sufficient for a large part of the direct image plates of our archives, but not so for the fine grained spectroscopic material nor for the very short focal length S40/50 films. Another limitation we identified is the amount of internally scattered light, and the different optical quality of the three RGB channels.
Therefore the initial activity of digitization was concentrated on images and on objective prism plates, leaving aside the spectra and the S40/50 material for which a small area 3200x6400 dpi scanner has been recently acquired.

In addition to *Scanfits*, another software tool has been developed at the Astrophysical Observatory of Catania, namely *AstroPlates*, a program that takes advantage of MS Visual Basic 6.0. It runs in any 32 bit Microsoft environment. *AstroPlates* acquires the images in TIFF format, either in B/W (14-16 bit/pixel) or RGB (14-16 bit/pixel for each channel). In the latter case the program is able to isolate a single channel. In the next step the images are transformed in FITS (with appropriate header), JPG or Bitmap format. It is also possible to record plate information in MDB database. An important characteristic of the program is the use of the powerful development environment IDL and its libraries by using the OCX component of IDL to communicate with it and to take advantage of its graphical and numerical capabilities. All functions of *AstroPlates* use the IDL libraries without activating the IDL environment. Therefore, it is not necessary that the user knows this language and its functions. During the acquisition phase it is possible to optimize the scanner parameters by a procedure taking into account the analysis of frequency histograms of grey levels in the plate.

Two different acquisition programs are therefore available, and the generic user can choose the one most suited to his taste or environment.

The approximate dimensions of the digitized files at 1600 dpi, 14 bit per dot, are: S67/92 cm plates (20x20 cm) 260 MB, 122 cm plates (9x12 cm) 70 MB, 182 cm plates (20x12 cm) 160 MB, S40/50 cm plates (10 cm diameter) 100 MB. The size of the files obviously poses a serious problem, both for storage and for distribution. After the initial period of test, when the files were saved to DVDs of 4.7 GB each, Network Attached Storage (NAS) units, capable to grow up to 1 TeraByte per unit, have already been implemented in Asiago and Roma.

## 3.1 Astrometry

The astrometric precision of the Asiago S67/92 digitized plates was tested in subfields of 1500x1500 pixels (1".55/px, 0.65x0.65 sq deg), with approximately 150 stars from the USNO Catalog (Monet et al. 2003) per field, finding a standard deviation in both coordinates of about 0"35. This precision is quite compatible with that obtained for instance by Barbieri et al. (1988) for the astrometry of Pluto using conventional measuring machines.

A further check was done in Torino as part of a research program dealing with the realization of the optical link to the quasi-inertial radio reference frame: approximately 400 plates taken around 1990, covering about 80 radio sources, were digitized using TO.CA.M.M, an original ASCORECORD device which was modified to be fully automated and housed at Cagliari Observatory in 1994.

As part of the standard calibration procedures, the geometrical/temporal stability of the measuring machine was monitored using rotated/repeated scans; tests have proved that TO.CA.M.M. routinely performes to better than 1 micron stability (Lattanzi et al. 2001, Bucciarelli et al. 2003).

In order to have an independent check of the astrometric properties of our scanners, we made a direct comparison between the *x,y* measurements on a particular plate digitized with TO.CA.M.M, our "reference catalog" in this context, and those obtained with the scanner Microtek Artixscan 1100, currently available at Torino. Setting the scanner at its maximum resolution - 1000 dpi - the effective pixel size of the image is ~ 25 microns, which we have reproduced with TO.CA.M.M. to maintain a comparable PSF sampling.

A 20x20 cm photographic plate of the series taken at the 38cm F/18 MORAIS refractor for the QSO program mentioned above was selected for this test.

An automatic match of the two lists of *x,y* measurements found 370 common objects over the entire plate; the vectors of differences in the two coordinates were minimized via a linear least squares fit, which gave an average positional error of 23 microns, corresponding to a disappointing 0.7 arcseconds. However, inspection of the individual residuals revealed the presence of systematic patterns in the residuals' behaviour as function of both *x* and *y* axes.

Therefore, as a first attempt to eliminate such systematics we selected a smaller region of the plate of about 5 cm (or 0.4 degrees) and recomputed a linear fit on the 27 matched objects.
In this case, we obtained a much better rms of 9.5 microns (or 0.28") and 4.2 (0.13") microns, in the *x* and *y* coordinates respectively.

These very encouraging results will be further investigated, but they give us confidence that the metric properties of our plates can be preserved by scanning them with a commercial scanner once the systematics are properly characterized

## 3.2 Photometry

Regarding the photometric accuracy attainable on the plates from the different telescopes, with a variety of focal lengths and emulsions, two procedures have been tested, both with good results. One procedure is the following: as no sensitometric spots are generally available for the plates, the plate transparency given by the scanner is transformed into Baker's photographic density, and then into flux assuming a standard density-intensity relation. The sequence of the various steps of the reduction procedures has been set up to work under IRAF and IDL environments.

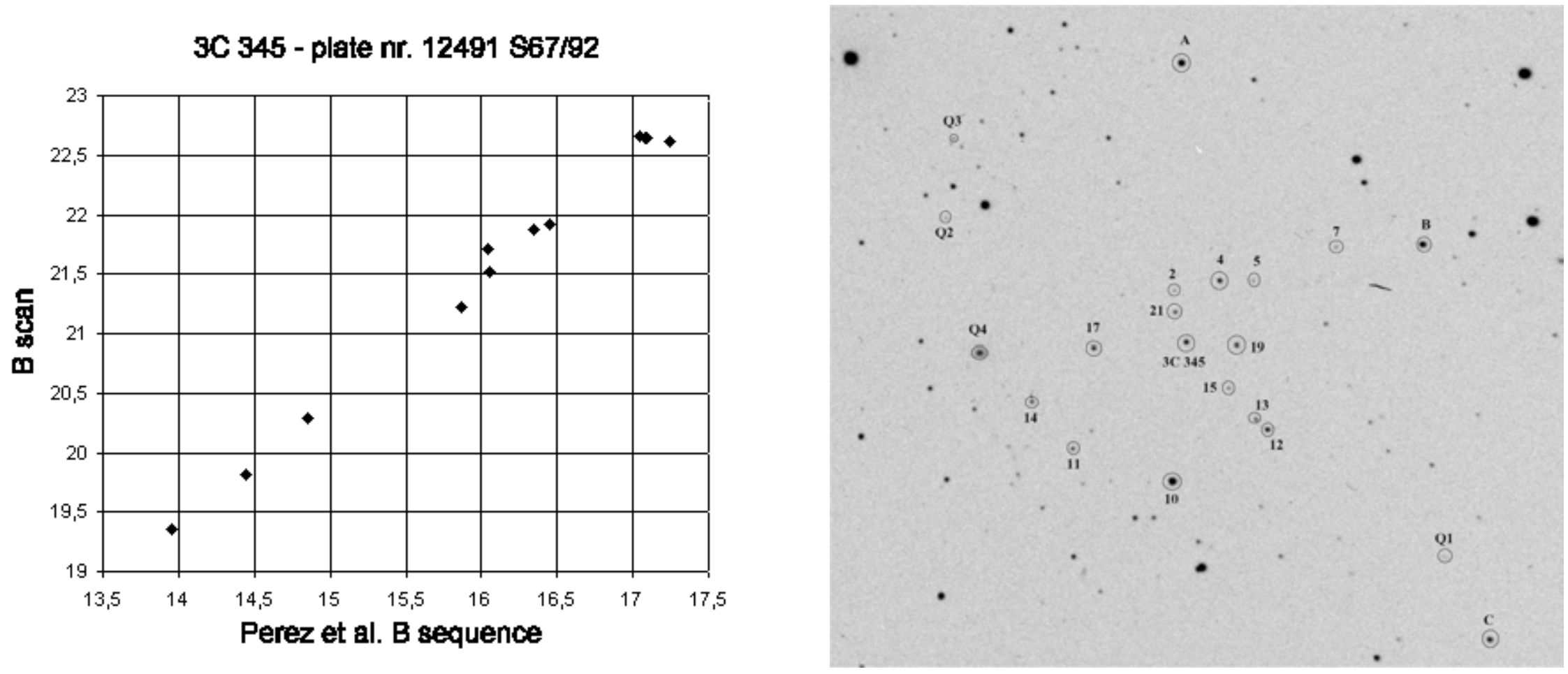

**Fig. 2 - Left: comparison between the B-scanner photometry and the B-CCD for the sequence of the field of 3C 345 published by Gonzales-Perez et al. and Angione (on the right, the digitised image of plate nr. 8031 S67/92).**

An example is given in Fig. 2, relative to the photometry of the Quasar 3C 345. The figure shows the comparison of the B magnitudes obtained with the scanner from an Asiago S67/92 Schmidt telescope with those of the CCD sequence published by Gonzalez-Perez et al., 2001. In this case aperture photometry was performed using IRAF/*apphot*. The linearity is fairly satisfactory, the standard deviation is 0.105 mag.

A second procedure has been implemented in collaboration with J. Johnson (U. Berkeley); it works under IDL using a PSF procedure based on the algorithms for the photographic emulsions presented by Stetson (1979). An example is shown in Fig. 3 where a blue plate of SA 57 obtained with the Asiago S67/92 cm (emulsion 103a-O + GG13) is confronted with Purgathofer's (1979) photoelectric B sequence. The method fits the magnitudes given by the scanner remarkably well in the range 8 to 18 mag. The rms from the linear fit is ±0.105 mag, dominated by two outlying stars.

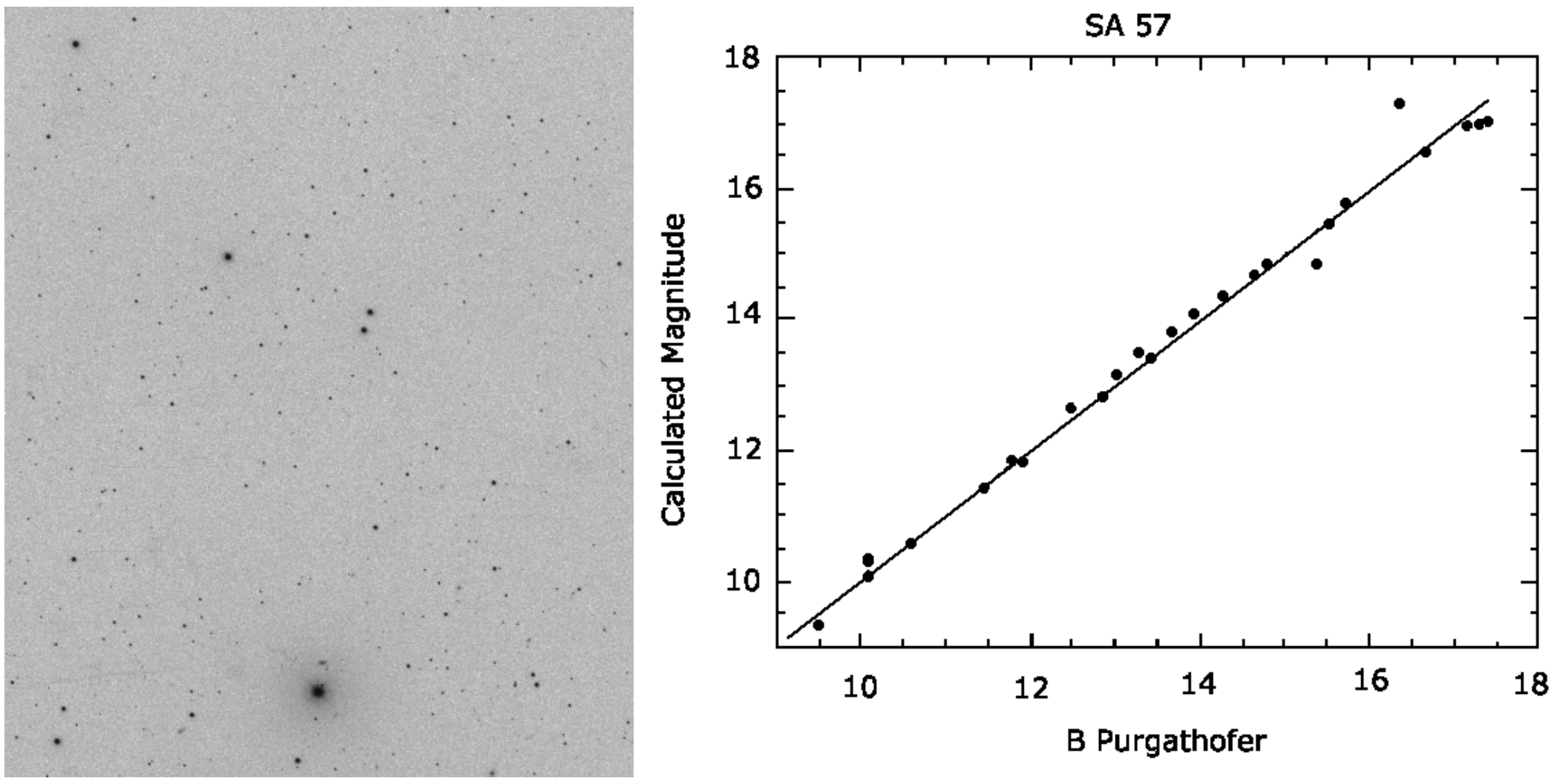

**Fig. 3 – B-scan photometry of SA 57 from an Asiago S67/92 Schmidt plate (left) vs Purgathofer photoelectric sequence using the PSF method.**

These tests confirm therefore that sufficiently accurate photometry can be done on the digitized images, permitting for instance to look for variable stars.

## 3.3 Objective prism plates

The Roma University team has taken up the task to digitize the objective prism material, given their long-standing interest on these topics (see Cassatella et al., 1973, 1975, Coluzzi et al., 1979). Some examples of spectral classification of stars on objective prism plates taken with the Campo Imperatore Schmidt telescope of the Roma Astronomical Observatory were presented by Barbieri et al. (2003a).

Tests on the digitization of objective prism plates of the Asiago 67/92 Schmidt are now under way. We have selected for this purpose two 103aF plates of the field of 3U 0352+30 (#453 and #507) taken in 1973 and 1976. Scans were made with the spectra aligned with the CCD detector, so that the length of the spectra is insensitive to possible irregularities in the scanning speed. After an approximate transformation of the recorded transparency into intensity (assuming gamma = 1.0), the plates were astrometrically calibrated using the position of the sharp red cut-off of the emulsion, and stars in the Tycho2 catalogue (Hog et al. 1998) as reference, using the procedure IRAF/ccmap. The precision is better than 2 arcsec.

Several spectra were then extracted with IRAF/apall and their tracing compared. Fig. 4 shows the spectra of the A0 star HD 281305 from two different scans of each plate. The plots have been shifted vertically for clarity, otherwise would be nearly indistinguishable, demonstrating that the scanner is well suited for the digitization of the Asiago objective prism plates. The overall shape of the spectra, dropping in the green, is dominated by the emulsion sensitivity. The Balmer series is well evident. The spectra from plate 507 have a better spectral resolution, probably due to a better atmospheric seeing or telescope focusing: the different plate quality is apparent also at visual inspection with a microscope.

We tried also a first automatic spectral classification on some stars of the field. We selected 11 stars with spectral classification from the HD catalogue (III/135A at CDS) as templates and classified automatically 22 other stars of the same catalogue. Our procedure first of all aligns all the spectra at the red cut-off of the emulsion sensitivity, then normalizes each spectrum to its continuum and finally makes the ratio of the unclassified spectrum with each template: the rms deviation is then computed and the spectral type is assumed to be that of the best fitting template. Our procedure gave to all the trial stars a spectral type within one step of our template grid, demonstrating that automatic spectral classification is feasible with the digitized plates. We are now working on improving the procedure, mainly testing if the template grid, based on bright (B<9) HD stars is suitable also for fainter stars down to the useful plate limit (B~13). A neural network approach (Philips et al. 2002) will also be explored.

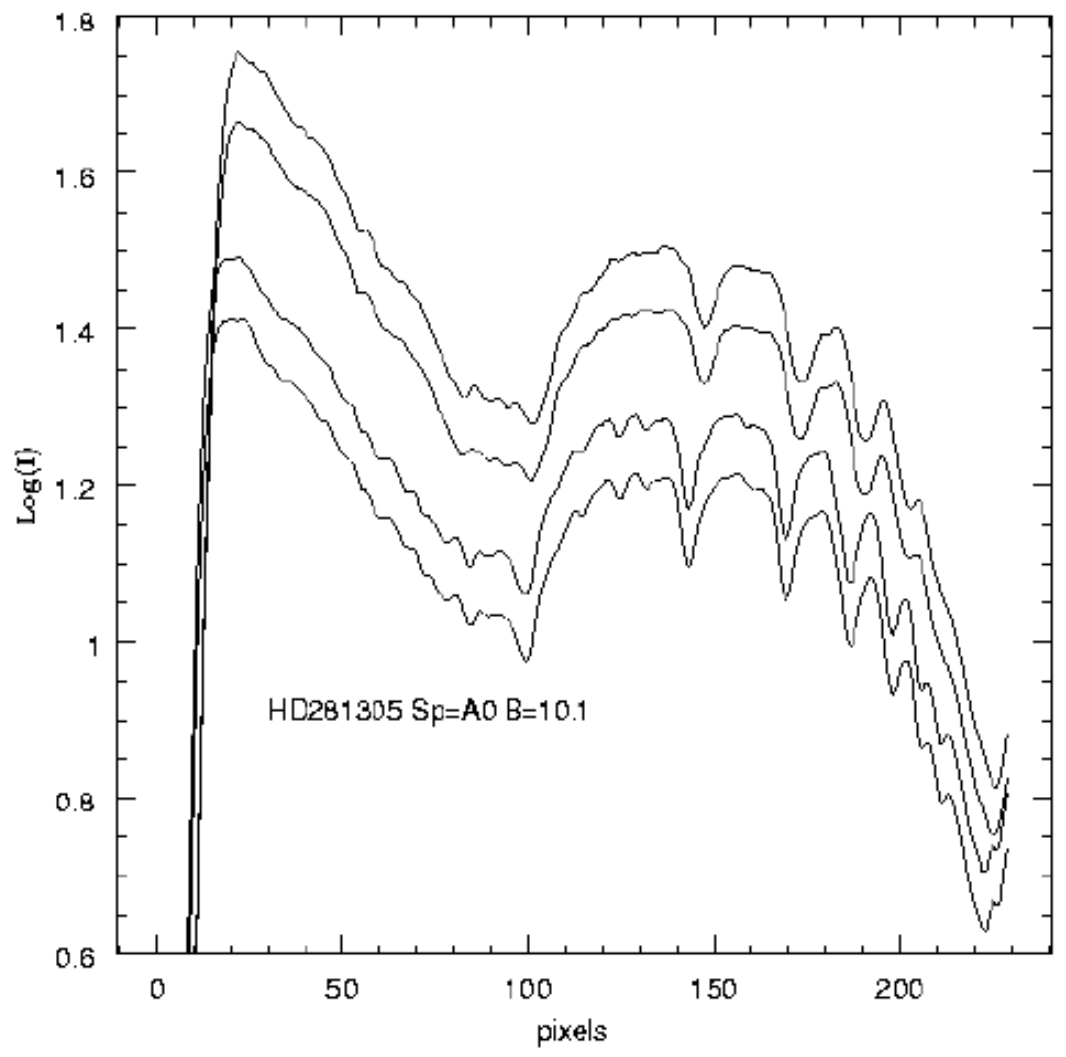

**Fig. 4 - Spectra of HD 281305, spectral type A0, from plate #453 (the two upper plots) and #507 (the two lower plots): red is at left.**

Both the astrometric calibration technique and the automatic extraction of the spectra based on the same model of scanner, are now being applied to the First Byurakan Survey (Mickaelian et al. 2004), in the frame of a collaboration between the Byurakan Observatory and the University of Rome presented in detail by Nesci et al. (2003).

Furthermore, we are analysing a set of objective prism plates taken with the Hamburg-Bergedorf 80/120/240 cm Schmidt telescope obtained during 1969-1975 by L. Kohoutek to follow the spectral evolution of the symbiotic nova HBV 475 (V1329 Cyg, see Fig. 5). The central part of the plates were digitized and the spectrophotometric calibration is being done, based on the techniques used by Baratta et al. (1976) for the analysis of peculiar stars in objective prism plates. Preliminary results have been presented by Rossi et al. (2003).

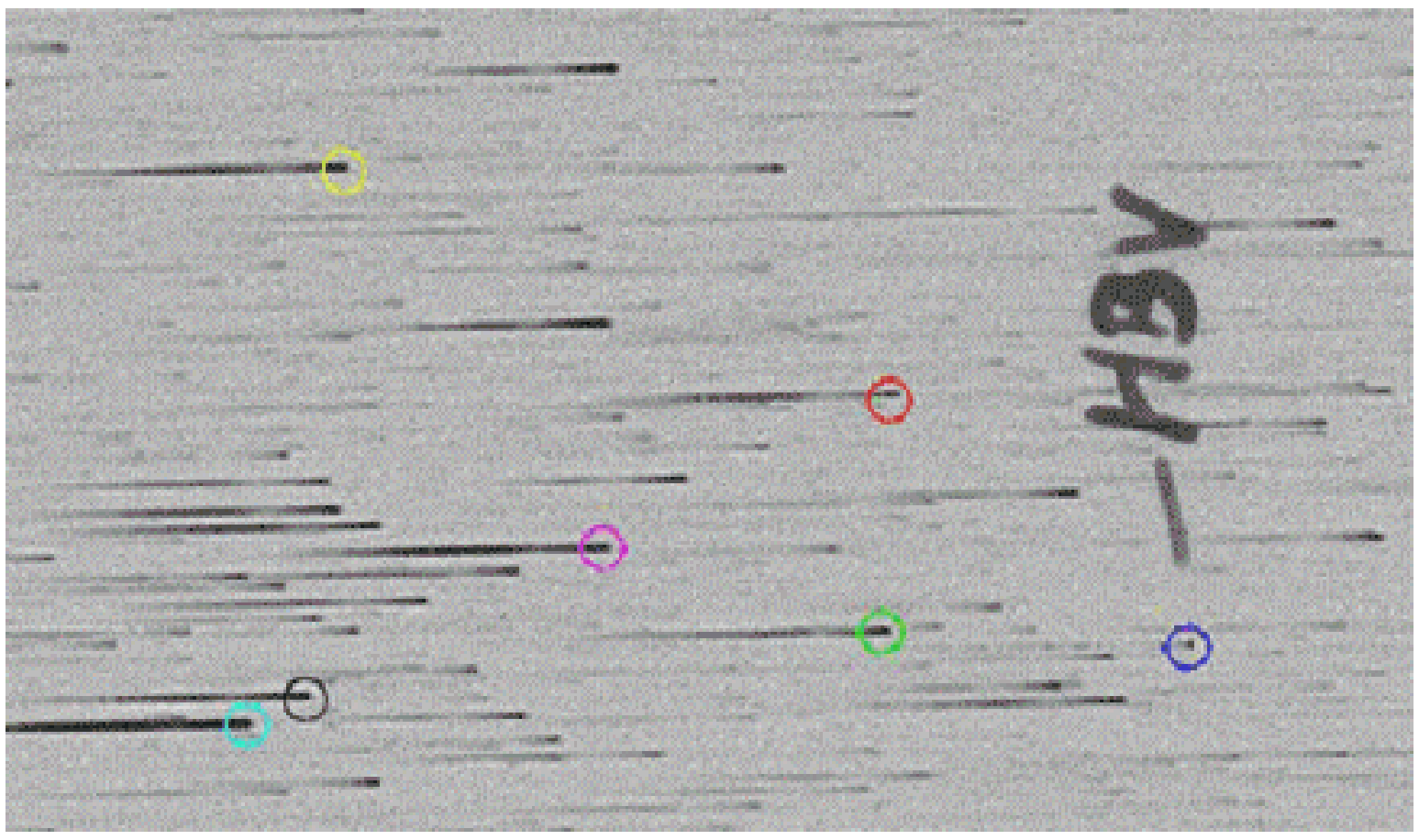

**Fig. 5 - The digitized central field of a Hamburg-Bergedorf objective-prism plate (unfiltered 103aE emulsion) near the exploded symbiotic star HBV 475 (Rossi et al. 2003). The spectra of HBV 475 and of a number of calibration stars are marked. North is to the left.**

## 3.4 Photometric calibration with new CCD sequences

The Roma Observatory has already started the acquisition of CCD calibration sequences in several sky fields. Observations are carried out at the 60/90/183 cm Schmidt telescope of the Campo Imperatore Observatory equipped with the 2048x2048 thinned and back illuminated CCD camera (ROSI) and a standard Johnson filters set. The aim of this work is to produce an algorithm to be used to linearize fluxes from the digitized plates, or to create a pipeline to directly modify the data files. The final solution will depend on which will make it easier for the final user to extract scientific information from the archive. A conversion to the standard Johnson filter set will also be attempted to homogenize actual measurements and data from the historical archive. The already observed fields in the M 31 region are shown in Fig. 6.

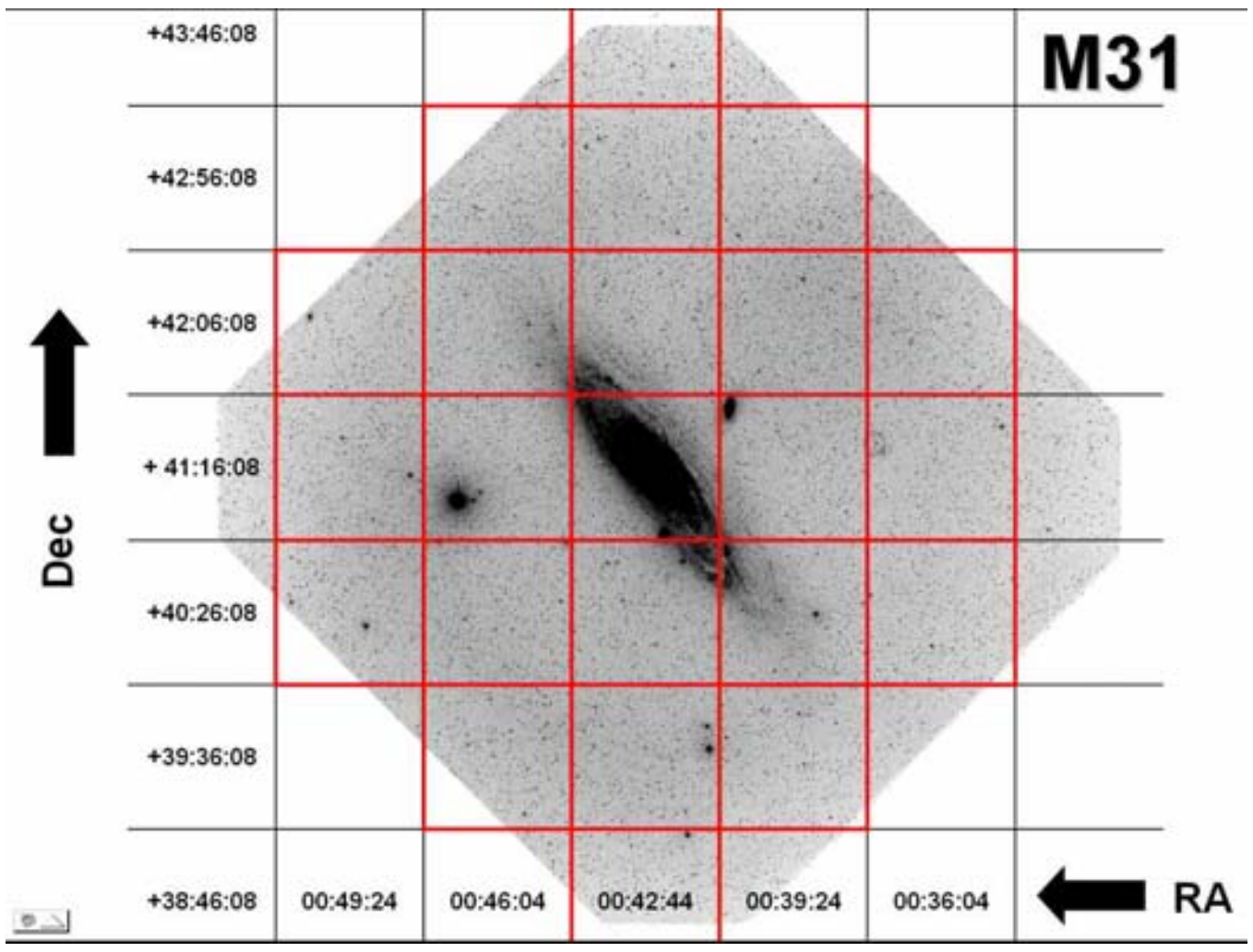

**Fig. 6 – The Campo Imperatore BVRI CCD fields superimposed on a digitized Asiago S67/92 plate of M 31.**

The Rome Observatory is also developing an automated pipeline to transform digital plate images to conform the astronomical standard for images in terms of orientation, fits header completeness and first approximation astrometry (to the arcsec level). This procedure has been verified to run on the Asiago Schmidt plates, running in background by converting all archive files without user support. The pipeline when finally released will include other telescopes support and possibly photometric calibration.

## 4 Status of the Digitization

More than 2000 plates have already been scanned, in the framework of different researches including:

- Study of stellar variability in M 31 and M 33: **all** the Asiago plates, including those originally used by Rosino and Bianchini (1973) for the search of Hubble-Sandage variables.
- Variability of QSOs and AGNs: **all** the Asiago plates of the field of 3C 345, extending the work of Barbieri et al. (1977) with still unpublished data.
- Search of stellar flares: 50 multiexposure plates of the Pleiades and NGC 7000 obtained with the Asiago S 67/92 cm Schmidt telescope, a joint project with the Sofia Sky Archive Data Center.
- All Comet Halley plates taken from the Catania Observatory during its 1910 passage.
- one hundred Asiago plates taken with the S 67/92 cm of the field of NGC 2264, on request by astronomers from Berkeley and Harvard
- spectral classification of stars and galaxies and spectrophotometric study of cataclysmic variable stars on 700 objective prism plates of the Vatican Observatory

## 5 Future Plans

We briefly sketch our plans for the immediate future:

- Complete the digitization of the log-books of all telescopes; in the site of the Asiago Observatory (http://www.pd.astro.it/Asiago/) we have implemented a section dedicated to photographic archives, where the already digitized logbooks are available in PDF format. For the S67/92 telescope, an on-line query page is available (see http://dipastro.pd.astro.it/~asiago/), yielding data from the main fields of the catalogue, and an example of the jepg preview for some of the already scanned plates. The query can be made by plate number, name or coordinates of the object.
- Implement by the beginning of next year the query on-line for all the Asiago telescopes log-books, translated from Italian into English, and by mid 2004 the possibility to display a high resolution jpeg preview and to directly download the selected Fits files.
- Make the FITS and JPG files accessible to the general user through the web. Start a call for proposals to the International Community, in order to selectively digitize those plates that give a maximum scientific return.
- The very nature of this project calls for its harmonisation with the concept of Virtual Observatory. The added value photographic plates can bring to the VO is to enhance the time axis of the VO multi-dimensional space. We plan therefore to coordinate our work with the Italian activities for the Datagrid and national Virtual Observatory (**DRACO:** http://wwwas.oat.ts.astro.it/draco/DRACO-home.htm). As such, the use of the standards defined within the working groups of the International VO Alliance (**IVOA:** http://www.ivoa.net) is envisaged, and data are planned to be eventually accessible to the community at large through the VO.

## Acknowledgments

We gratefully acknowledge the support of many collaborators: S. Magrin, V. Mezzalira, A. Migliorini and G. Umbriaco (Dept. of Astronomy, Padova University), E. Massone (Torino Observatory) S. Sclavi (student of the Roma University La Sapienza), E. Catinoto (Catania Observatory). John Johnson (Berkeley U.) kindly provided his software routines for converting the photographic data into linear units.

**Keywords:** Photographic Archives, Digitization of Astronomical Plates, Objective Prism Plates, Photometric Calibration